\documentclass[prd,preprint,superscriptaddress,amsmath,amssymb,nofootinbib]{revtex4}
\usepackage{graphicx}
\usepackage{dcolumn}
\usepackage{bm}
\usepackage{amssymb}
\usepackage{amsmath}
\usepackage{epsfig}    
\usepackage{color}
\usepackage{slashed}
\usepackage{hhline}

\def\be{\begin{equation}}
\def\ee{\end{equation}}
\newcommand{\bea}{\begin{eqnarray}}
\newcommand{\eea}{\end{eqnarray}}
\newcommand{\nn}{\nonumber}

\begin{document}

{\begin{flushright}{APCTP Pre2022 - 018}\end{flushright}}

\title{Muon $g-2$ with $SU(2)_L$ multiplets}

\author{Takaaki Nomura}
\email{nomura@scu.edu.cn}
\affiliation{College of Physics, Sichuan University, Chengdu 610065, China}

\author{Hiroshi Okada}
\email{hiroshi3okada@htu.edu.cn}
\affiliation{Department of Physics, Henan Normal University, Xinxiang 453007, China}
\affiliation{Asia Pacific Center for Theoretical Physics (APCTP) - Headquarters San 31, Hyoja-dong,
Nam-gu, Pohang 790-784, Korea}
\affiliation{Department of Physics, Pohang University of Science and Technology, Pohang 37673, Republic of Korea}

\date{\today}

\begin{abstract}
We propose a simple model to obtain sizable muon anomalous magnetic dipole moment (muon $g-2$) introducing several $SU(2)_L$ multiplet fields without any additional symmetries. The neutrino mass matrix is simply induced via type-II seesaw scenario in terms of $SU(2)_L$ triplet Higgs with $U(1)_Y$ hypercharge 1. In addition, we introduce an $SU(2)_L$ quartet vector-like fermion with $1/2$ hypercharge and scalar with $3/2$ hypercharge. The quartet fermion plays a crucial role in explaining muon $g-2$ causing the chiral flip inside a loop diagram with mixing between triplet and quartet scalar bosons via the standard model Higgs. We show numerical analysis and search for allowed region in our parameter space, and demonstrate the collider physics.
 \end{abstract}
\maketitle

\section{Introductions}
Even after discovery of the standard model (SM) Higgs, we have to resolve several issues such as non-zero neutrino masses and muon anomalous magnetic dipole moment (muon $g-2$) that would indicate necessity of beyond the SM.
New results on the muon $g-2$ are reported by the E989 collaboration at Fermilab \cite{Muong-2:2021ojo,Muong-2:2023cdq}: 
\begin{align}
 a^{\rm FNAL}_\mu =116592055(24) \times 10^{-11}. 
\label{exp_dmu}
\end{align}
Furthermore, combined result with the previous BNL, suggests that the muon $g-2$ deviates from the SM prediction by 5.1$\sigma$ level~\cite{Muong-2:2021ojo, Muong-2:2023cdq, Aoyama:2012wk,Aoyama:2019ryr,Czarnecki:2002nt,Gnendiger:2013pva,Davier:2017zfy,Keshavarzi:2018mgv,Colangelo:2018mtw,Hoferichter:2019mqg,Davier:2019can,Keshavarzi:2019abf,Kurz:2014wya,Melnikov:2003xd,Masjuan:2017tvw,Colangelo:2017fiz,Hoferichter:2018kwz,Gerardin:2019vio,Bijnens:2019ghy,Colangelo:2019uex,Blum:2019ugy,Colangelo:2014qya,Hagiwara:2011af}:
\begin{align}
\Delta a^{\rm new}_\mu = (24.9\pm 4.9)\times 10^{-10}. 
\label{exp_dmu}
\end{align}
Although results on the hadron vacuum polarization (HVP), estimated by recent lattice calculations~\cite{Borsanyi:2020mff,Alexandrou:2022amy,Ce:2022kxy}, may weaken the necessity of a new physics effect, it is shown in refs.~\cite{Crivellin:2020zul,deRafael:2020uif,Keshavarzi:2020bfy}~\footnote{The effect in modifying HVP for muon $g-2$ and electroweak precision test are also discussed previously in ref.~\cite{Passera:2008jk}.} that the lattice results imply new tensions with the HVP extracted from $e^+ e^-$ data and the global fits to the electroweak precision observables. 
However, we note that such tensions only occur at large $q^2$ region, while a shift in the $e^+e^-$ hadronic cross section for momentum transfer below 1 GeV (e.g. from $e^+e^- \to \pi^+ \pi^-$) does not give such issue.
In addition the CMD-3 collaboration \cite{CMD-3:2023alj}  released results on the 
cross section of $e^+ e^- \to \pi^+ \pi^-$ that disagree at the $(2.5-5) \sigma$ level with all previous measurements that weakens deviation of muon $g-2$.
Thus it is controversial about the origin of the anomaly and further experimental/theoretical explorations are needed.
If muon $g-2$ suggests new physics, we expect new particles and interactions. 
To explain the sizable muon $g-2$ with natural manner by Yukawa couplings,~\footnote{If one explains it via new gauge sector such as $U(1)_{L_\mu-L_\tau}$, chiral flip is not needed but narrow region as for the gauge coupling and its mass~\cite{Altmannshofer:2014pba}.} we would need one-loop contributions with chiral flip by heavy fermion mass inside a loop diagram~\cite{Lindner:2016bgg, Athron:2021iuf,Guedes:2022cfy,Crivellin:2021rbq}.
Otherwise, the Yukawa couplings would exceed perturbation limit or too light mediator masses are required.

Simple ways to extend the SM to resolve these issues are introduction of new fields that are $SU(2)_L$ multiplets~\cite{Nomura:2018lsx, Nomura:2018cfu, Nomura:2018ibs, Nomura:2018cle, Nomura:2017abu,Nomura:2016jnl, Nomura:2016dnf,Cai:2017jrq,Anamiati:2018cuq,Guedes:2022cfy,Calibbi:2018rzv,Baek:2016kud,Chen:2020jvl}.
For example, neutrino masses can be induced by adding Higgs triplet with hypercharge $1$ that is known as type-II seesaw mechanism~\cite{Magg:1980ut,Lazarides:1980nt,Schechter:1980gr,Cheng:1980qt,Mohapatra:1980yp,Bilenky:1980cx}.
We can also expect that sizable contribution to muon $g-2$ is obtained by adding a vector-like fermion multiplet in addition to a scalar multiplet 
where chiral flip occurs inside a loop picking up vector-like fermion mass.
Also multiple electric charge of components in large multiplets can enhance muon $g-2$ value.
In addition to explaining muon $g-2$ anomaly and neutrino masses, large $SU(2)_L$ multiplet fields would induce interesting signatures at collider experiments as it contains multiply-charged particles.

In this paper, we explain the sizable muon $g-2$ via $SU(2)_L$ multiplet fields without any additional symmetries.
More concretely,  we add an $SU(2)_L$ quartet  vector-like fermion $\psi$ with $1/2$ hypercharge, one triplet Higgs $\Delta$ with $1$ hypercharge, and one quartet scalar $H_4$ with $3/2$ hypercharge. The quartet fermion plays an crucial role in explaining the sizable muon $g-2$ causing the chiral flip in terms of its mass term as well as through mixing between triplet and quartet bosons.
In addition, the neutrino mass matrix is simply induced via type-II scenario via the Yukawa interactions between the lepton doublet and triplet Higgs field.
The choice of $SU(2)_L$ quartet  vector-like fermion $\psi$ is suitable to obtain Yukawa interactions of $\bar L_L \Delta^\dagger \psi_R$ for muon $g-2$; 
we can have similar term with $SU(2)_L$ doublet  vector-like fermion but we also have undesired term of $\bar L_L \psi_L^c$ inducing unnecessary mixing between the SM lepton and the vector like fermion. 
Then we need $H_4$ to make the Yukawa term $\bar \psi_L h_4 e_R$ to get chiral flip in the loop diagram inducing muon $g-2$.
Note also that we should consider constraints on vacuum expectation values (VEVs) of $SU(2)_L$ multiplet scalar fields since it deviate $\rho$-parameter from $1$.
After formulating our model, we show numerical analysis and search for allowed region in our parameter space, and discuss the collider physics focusing on productions of multiply-charged particles in the model.

This paper is organized as follows.
In Sec. II, we introduce our model and formulate the Yukawa sector and Higgs sector, oblique $\rho$ parameter,
neutral fermion masses including the active neutrino masses, lepton flavor violations (LFVs), and muon $g-2$.
In Sec. III, we show numerical analysis of muon $g-2$ and discuss collider physics.
Finally we devote the summary of our results and the conclusion.

\section{Model setup and Constraints}
\begin{table}[t!]
\begin{tabular}{|c||c|c|c||c|c|c|}\hline\hline  
& ~$L_{L_i}$~& ~$e_{R_i}$~& ~$\psi$~ & ~$H$~& ~$\Delta$~& ~$H_4$~ \\\hline\hline 
$SU(2)_L$   & $\bm{2}$  & $\bm{1}$  & $\bm{4}$ & $\bm{2}$ & $\bm{3}$  & $\bm{4}$     \\\hline 
$U(1)_Y$    & $-\frac12$  & $-1$ & $\frac12$ & ${\frac12}$ & $1$  & $\frac32$ \\\hline
\end{tabular}
\caption{Charge assignments of the our lepton and scalar fields
under $SU(2)_L\times U(1)_Y$, where the lower index $i$ is the number of family that runs over $1-3$,
all of them are singlet under $SU(3)_C$, and the quark sector is the same as the SM one.}\label{tab:1}
\end{table}

In this section we introduce our model.
As for the fermion sector, we introduce one family of  vector-like fermion $\psi$ with $(4,{1/2})$
where each of content in parentheses represents the charge assignment of the  SM gauge groups ($SU(2)_L, U(1)_Y$), hereafter.
As for the scalar sector, we add a triplet scalar field $\Delta$ with $(3,1)$ which realizes type-II seesaw mechanism
 and a quartet scalar field $H_4$  with $(4,{3/2})$, where SM-like Higgs field is denoted as $H$.
Here we write components of multiplets as
\begin{align}
& H = (h^+, \tilde h^0) \\
& \Delta = \begin{pmatrix} \frac{\delta^+}{\sqrt2} & \delta^{++} \\ \delta^0 & -\frac{\delta^+}{\sqrt2}  \end{pmatrix},\label{eq:H3}  \\
& H_4 = ( \phi^{+++}_4,\phi_4^{++}, \phi_4^{+}, \phi_4^{0})^T,  \label{eq:H4} \\
& \psi_{L(R)} = ( \psi^{++}, \psi^{+},\psi^{0}, \psi^{-})^T_{L(R)} \label{eq:psiLR},
\end{align}
where $\tilde h^0 = \frac1{\sqrt2} (h^0 + v_h + i G^0)$ and the triplet can be also written by $H_3 =( \delta^{0}, \delta^{+}, \delta^{++})^T$.
Neutral components of scalar fields develop VEVs denoted by $ \{ \langle H\rangle, \langle \Delta \rangle, \langle H_4 \rangle \} \equiv \{v_h, v_\Delta, v_4  \}/\sqrt2$ which induce the spontaneous electroweak symmetry breaking.
All the field contents and their assignments are summarized in Table~\ref{tab:1}, where the quark sector is exactly the same as the SM.
The renormalizable lepton Yukawa Lagrangian under these symmetries is given by
\begin{align}
-{\cal L_\ell}
& =  y_{\ell_{ii}} \overline{L_{L_i}} H e_{R_i} 
+ y_{\nu_{ij}} \overline{L_{L_i}} \Delta^\dagger L^c_{L_j}  
+  y_{D_{i}} [ \overline{L_{L_i}} \Delta^\dagger \psi_{R} ]  
+  f_{i} [\overline{\psi_{L}}  H_4 e_{R_i}]\nn\\
&
+  g_{L} [\overline{\psi^{c}_{L}}  \Delta^\dagger \psi_{L}]
+  g_{R} [\overline{ \psi^{c}_{R}}  \Delta^\dagger \psi_{R}]
+M_{\psi_{}} \overline{ \psi_{L}} \psi_{R}
+ {\rm h.c.}, \label{Eq:yuk}
\end{align}
where we implicitly symbolize the gauge invariant contracts of $SU(2)_L$ index as bracket [$\cdots$] hereafter,
indices $(i,j)=1$-$3$ are the number of families, $y_\ell$ is assumed to be  diagonal matrix with real parameters
without loss of generality.
Then, the mass eigenvalues of charged-lepton are defined by $m_\ell=y_\ell v_h/\sqrt2={\rm Diag}(m_e, m_\mu, m_\tau)$. 
%
In our model scalar potential is written by
\begin{align}
V = & - \mu_H^2 H^\dagger H + \mu_\Delta^2 {\rm Tr}[\Delta^\dagger \Delta] + \mu_{H_4}^2 H^\dagger_4 H_4 + \lambda_H (H^\dagger H)^2  \nonumber \\
& + (\text{trivial quartet terms including $\Delta$ and $H_4$}) + V_{\rm non-trivial},
\end{align}
where we omit details of trivial quartet terms with $\Delta$ and $H_4$ for simplicity and assume their couplings are small.
The non-trivial scalar potential is given by
\begin{align}
V_{\rm non-trivial} = \mu_1 [H^\dag \Delta^\dag H_4+{\rm h.c.}] +\mu_2 [H^T \Delta^\dag H]
+ \sum_i \lambda_{H_4 H}^i [H_4^\dag H H H]_i + {\rm h.c.},
\label{Eq:potential}
\end{align}
where $\mu_1$ plays a crucial role in inducing the muon $g-2$ as can be seen later.

 Here we discuss the advantage of choosing $SU(2)$ quartet for scalar and fermion.
Firstly we would like to have interaction terms of $\overline{L_L} \Delta^\dagger \psi_R$, $\overline{\psi_L} H_4 e_R$ and $H^\dagger \Delta^\dagger H_4$ to get chirality flip enhancement for muon $g-2$ while realizing neutrino mass via type-II seesaw.
In fact we can generalize $SU(2)_L$ representation of $\psi$ and $H_4$ to be $\bf{N}$ if it satisfies $\bf{N} \times \bf{3} \times \bf{2} \supset \bf{1}$ to get above terms. 
The minimal choice is  $\bf{N}=2$ writhing new scalar and fermion as $H_2$ and $\psi'$, but this case induces non-desired terms such as $\overline{L_L} \psi'^c_L$ and $\overline{\psi'^c_R} H_2 e_R$.
These terms would induce non-negligible mixing between the SM charged leptons and exotic charged fermions.
We thus choose $\bf{N} = \bf{4}$ to avoid these unnecessary terms.
The choice of larger multiplet also enhances muon $g-2$ as we have more contributions from components in multiplets.
In addition, the choice of $\bf{N} = \bf{4}$ induces interesting phenomenology at the collider experiments as it provides multiply-charged particles inside a multiplet.

\subsection{VEVs of scalar fields and $\rho$-parameter}

Non-zero VEVs of scalar fields are obtained by solving the stationary conditions 
\begin{equation}
\frac{\partial V}{\partial v_h} = \frac{\partial V}{\partial v_\Delta} = \frac{\partial V}{\partial v_4} = 0.
\end{equation}
Here we explicitly write the first two terms of Eq.~\eqref{Eq:potential} by
\begin{align}
\frac{\mu_1 }{3 \sqrt{2}} (v_h+h_0) (\sqrt{3} \phi^0_4 \delta^{0*} + \sqrt{6} \phi^+_4 \delta^- + 3 \phi^{++} \delta^{--}) - \frac{1}{\sqrt2} \mu_2 \delta^{0*} (h^0 + v_h)^2 +c.c. \ ,
\end{align}
where we consider it in unitary gauge.
Assuming $v_4, v_\Delta \ll v_h$ and small couplings for trivial quartet couplings, we obtain the VEVs approximately as
\begin{equation}
v_h \simeq \sqrt{\frac{\mu_H^2}{\lambda_H}}, \quad 
v_\Delta \simeq \frac{1}{\mu_\Delta^2} \left( \frac13 \sqrt{\frac32} \mu_1 v_4 v_h +  \mu_2 v_h^2 \right), \quad
v_4 \simeq \frac{1}{3}   \sqrt{\frac32} \frac{\mu_1 v_\Delta v_h}{\mu_{H_4}^2}. 
\label{eq:VEVs}
\end{equation}
Thus small values of $v_\Delta$ and $v_4$ are naturally obtained when mass parameters $\mu_\Delta$ and $\mu_{H_4}$ are larger than electroweak scale.

The electroweak $\rho$ parameter deviates from unity due to the nonzero values of $v_\Delta$ and $v_4$ at the tree level as follows:
\begin{align}
\rho=\frac{v_h^2+2 v_\Delta^2 + 6 v_4^2}{v_h^2+4 v_\Delta^2 + 9 v_4^2},
\end{align}
where the VEVs satisfy the relation $v \equiv \sqrt{v_h^2+v_\Delta^2+v_4^2} \simeq 246$ GeV. 
Here we consider current constraint on $\rho$ parameter; $\rho = 1.00038 \pm 0.00020$~\cite{ParticleDataGroup:2020ssz}.
If we take $v_X \equiv v_\Delta = v_4$ the upper bound of $v_X$ is 
\begin{equation}
v_X \lesssim 1.55 \ {\rm GeV},
\end{equation}
when we require $\rho$ to be within 2$\sigma$ level. 
In our analysis we choose $v_\Delta \sim v_4 \sim 1$ GeV for simplicity~\footnote{ Here we just choose the value around the upper limit from $\rho$-parameter. In this case we should require tiny Yukawa coupling ($y_\nu < 10^{-9}$) for neutrino mass generated by type-II seesaw mechanism. We can get smaller $v_\Delta$ by adjusting the parameters in the scalar potential.}.
 Note that in the model smallness of VEVs of triplet and quartet scalars can be obtained by large values of $\mu_\Delta$ and $\mu_4$ as in the Eq.~\eqref{eq:VEVs}.
The smallness of VEVs can be kept as long as these parameters are larger than cubic coupling $\mu_{1,2}$ even if there is a radiative correction.
Although higher order radiative correction would affect the VEVs we can tune these free parameters to make VEVs small in general.

Finally we briefly discuss vacuum stability of the scalar potential. 
In the model we choose scales of $\mu_\Delta$ and $\mu_{H_4}$ are much larger than VEVs of scalar fields. 
Then we obtain 
\begin{align}
& \frac{\partial^2 V}{\partial \delta^{0} \partial \delta^0} \simeq \frac{\partial^2 V}{\partial \delta^{+} \partial \delta^+} \simeq \frac{\partial^2 V}{\partial \delta^{++} \partial \delta^{++}} \simeq  \mu^2_\Delta, \nonumber \\
& \frac{\partial^2 V}{\partial \phi_4^{0} \partial \phi_4^{0}} \simeq \frac{\partial^2 V}{\partial \phi_4^{+} \partial \phi_4^{+}} \simeq  \frac{\partial^2 V}{\partial \phi_4^{++} \partial \phi_4^{++}} \simeq 
 \frac{\partial^2 V}{\partial \phi_4^{+++} \partial \phi_4^{+++}} \simeq \mu_{H_4}^2,
 \end{align}
and the other second derivatives of the potential are much smaller.
This condition will be kept after diagonalizing mass matrices of scalar bosons and the original components are approximately mass eigenstates since off-diagonal components of mass matrices are much smaller than diagonal components.
Thus the stability of the vacuum can be guaranteed by the positive values of $\mu_\Delta^2$ and $\mu_{H_4}^2$ in the model.
Also we assume all the coupling constants associated with quartic terms in the potential to be positive for requiring the absence of directions in scalar field space for which the potential is not bounded from below


\subsection{Masses of new particles}

The scalars and fermions with large $SU(2)_L$ multiplet provide exotic charged particles.
 The mass terms of  $H_4$, $\Delta$ and $\psi$ are approximately given by 
 \begin{equation}
 \mathcal{L}_M =  \mu_\Delta^2 {\rm Tr}[\Delta^\dagger \Delta] + \mu_{H_4}^2 H^\dagger_4 H_4 + \frac{\mu_1 v_h}{3} (\sqrt{3} \phi^0_4 \delta^{0*} + \sqrt{6} \phi^+_4 \delta^- + 3 \phi^{++} \delta^{--} + c.c.) + M_\psi \bar \psi \psi,
 \end{equation}
 where we ignored contributions from quartet terms in the scalar potential assuming they are small enough.
 Thus components in $\psi$ have degenerate mass $M_\psi$ where small mass shift appears at loop level~\cite{Cirelli:2005uq} but we ignore it in our analysis below.
 The triply charged scalar mass is given by $m_{H^{+++}} = \mu_{H_4}$ while 
%
we have $\delta^\pm-\phi^\pm_4$,  $\delta^{\pm\pm}-\phi^{\pm\pm}_4$, and $\delta^0-\phi^0_4$ mixings through $\mu_1$ term 
that lead to sizable muon $g-2$ as we discuss below.
We write mass eigenstates and mixings as follows:
\begin{align}
&\begin{pmatrix} \delta^\pm \\ \phi^\pm_4 \end{pmatrix} 
= 
\begin{pmatrix} c_\alpha & s_\alpha \\ - s_\alpha & c_\alpha \end{pmatrix} 
\begin{pmatrix}  H^\pm_1 \\ H^\pm_2 \end{pmatrix}, \label{eq:scalar-mass-fields1}  \\ 
&\begin{pmatrix} \delta^{\pm\pm} \\ \phi^{\pm\pm}_4 \end{pmatrix} 
=
\begin{pmatrix} c_\beta & s_\beta \\ - s_\beta & c_\beta \end{pmatrix} 
\begin{pmatrix}  H^{\pm \pm}_1 \\ H^{\pm \pm}_2 \end{pmatrix}, \label{eq:scalar-mass-fields2} \\ 
&\begin{pmatrix} \delta^{0} \\ \phi^{0}_4 \end{pmatrix} 
=
\begin{pmatrix} c_\gamma & s_\gamma \\ - s_\gamma & c_\gamma \end{pmatrix} 
\begin{pmatrix}  H^0_1 \\ H^0_2 \end{pmatrix},
\label{eq:scalar-mass-fields3}
\end{align}
where $c_{a},s_{a}$ are respectively short-hand notation of $\cos a,\sin a$ with $a\equiv(\alpha,\beta,\gamma)$.
The mass eigenvalues and mixing angles are given by
\begin{align}
& m^2_{\{H_1^{+}, H_1^{++}, H_1^{0} \}} = \frac12 (\mu_{H_4}^2 + \mu_\Delta^2) - \frac12 \sqrt{(\mu_{H_4}^2 - \mu_\Delta^2)^2 + 4 \Delta M^4_{\{+, ++, 0\}} }, \\
& m^2_{\{H_2^{+}, H_2^{++}, H_2^{0} \}} = \frac12 (\mu_{H_4}^2 + \mu_\Delta^2) + \frac12 \sqrt{(\mu_{H_4}^2 - \mu_\Delta^2)^2 + 4 \Delta M^4_{\{+, ++, 0 \} }} ,  \\
& \tan (2 \{\alpha, \beta, \gamma \}) = \frac{2 \Delta M^2_{\{+, ++, 0 \} }}{ \mu_{\Delta}^2 - \mu_{H_4}^2}, \\
& \Delta M^2_{\{+, ++, 0 \} }  = \left\{ \frac{\sqrt{3} \mu_1 v_h}{3}, \frac{\sqrt{6} \mu_1 v_h}{3}, \mu_1 v_h  \right\}.
\end{align}
Notice here that we neglect the mixing between the SM Higgs and other neutral scalar bosons choosing related parameters to be sufficiently small,
 and we do not discuss experimental constraint related to the SM Higgs boson assuming its couplings are the SM like.
 For example, the mixing between $\delta^0$ and $h^0$ is estimated by $\mu_1 v_4/\mu_\Delta^2$. The mixing angle is around $2 \times 10^{-3}$ if $v_\Delta = 1$ GeV, $\mu_1 = \mu_{H_4} = 1.2 M_\psi$, $\mu_\Delta = 0.8 M_\psi$ and  $M_\psi = 1$ TeV that is maximal angle in our numerical analysis. The mixing angle $2 \times 10^{-3}$ is sufficiently small to satisfy experimental constraints regarding Higgs boson measurement.

\subsection{Neutral fermion masses}

After the spontaneous symmetry breaking, neutral fermion mass matrix in basis of $\Psi^0_L\equiv (\nu_L^c, \psi_R,\psi_L^{c})^T$ is given by
\begin{align}
M_N
&=
\left[\begin{array}{ccc}
m_\nu^{(II)} & m_D & 0  \\ 
m_D^T & m_R & M_\psi \\ 
0  & M_\psi & m_L \\ 
\end{array}\right],
\end{align}
where $m_\nu^{(II)}\equiv y_\nu v_\Delta$, $m_D\equiv {y_D v_\Delta}/\sqrt3$, $m_R\equiv {2 g_R v_\Delta}/3$, and $m_L\equiv {2 g_L v_\Delta}/3$.
Achieving the block diagonalizing, we find the active neutrino mass matrix:
\begin{align}
m_{\nu_{}}\approx m_{\nu_{}}^{(II)} + \frac{m_D m_D^T m_L}{M^2_\psi}.
\end{align}
The second term in the above equation corresponds to inverse seesaw, but its matrix rank is one.
Thus, we simply expect that the neutrino oscillation data is dominantly described by the first term $m_{\nu_{}}^{(II)}$. 
Notice here that we need the following constraint to achieve $m_{\nu_{}}\approx m_{\nu_{}}^{(II)}$;
\begin{align}
\frac{m_D m_D^T m_L}{M^2_\psi} \ll 0.1\ {\rm eV}.
\end{align}
 It can be realized by requiring $m_L$ to be small; for example, if $m_D \sim 1$ Gev and $M_\psi = 1$ TeV we choose $m_L \ll 10^{-3}$ GeV. 
It is possible to make $m_L$ small since $g_L$ is free parameter.
Here, we also assume $m_D$ to be negligibly tiny compared to $M_\psi$ in order to evade the mixing between the SM charged-leptons  and the exotic charged fermions. In this case, there is no mixing between the active neutrinos and heavier neutral fermions also.
Thus, the heavier neutral mass eigenvalues diag[$D_1,D_2$] are given by unitary matrix $V_N$ as
$D =V_N M_N V_N^T$ where 
\begin{align}
M_N&=
\left[\begin{array}{cc}
m & M_\psi \\ 
M_\psi & m \\ 
\end{array}\right],\\
D_1&=M_\psi-m,\ D_2=M_\psi+m,\\
V_N&=\frac{1}{\sqrt2}
\left[\begin{array}{cc}
i & 0 \\ 
0 & 1 \\ 
\end{array}\right]
\left[\begin{array}{cc}
1 & -1 \\ 
1 & 1 \\ 
\end{array}\right].
\end{align}
Here we assume $m\equiv m_R=m_L$ for simplicity.

The active neutrino mass matrix is diagonalized by $D_\nu =U_{\rm MNS} m_\nu U_{\rm MNS}^T$  where $U_{\rm MNS}$ is Maki Nakagawa Sakata mixing matrix~\cite{ParticleDataGroup:2020ssz}.
It suggests that we simply parametrize $y_\nu$ as follows:
\begin{align}
y_\nu=\frac1{v_\Delta} U_{\rm MNS}^\dag D_\nu U_{\rm MNS}^T.
 \end{align}
Basically we can realize neutrino mass and mixing tuning Yukawa couplings $y_{\nu_{ij}}$ same as the type-II seesaw mechanism.
Thus we do not discuss neutrino masses further in this paper.

\subsection{ Lepton flavor violations(LFVs) and muon $g-2$}
\begin{figure}[tb]
\begin{center}
\includegraphics[width=10.0cm]{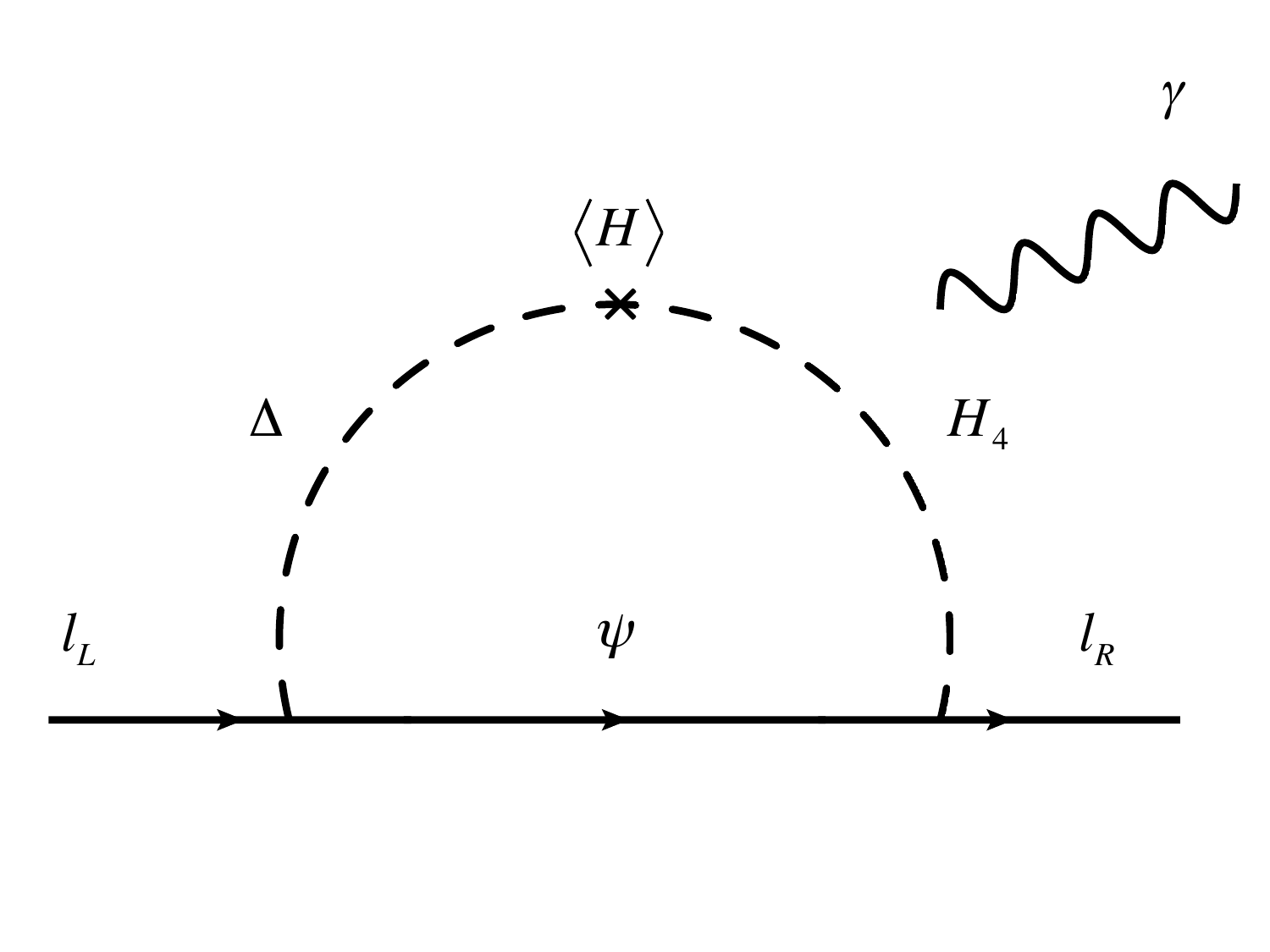}
\caption{Feynman diagram to generate muon $g-2$ and $\ell \to \ell' \gamma$ processes.}
\label{fig:diagram}
\end{center}\end{figure}

In our model LFV processes and muon $g-2$ are induced from Yukawa interactions associated with couplings $\{ y_D,\ f \}$.
The relevant terms are explicitly written by
\begin{align}
& f_i [\overline{\psi_L} H_4 e_{R_i}] + y_{D_i} [\overline{L_{L_i}} \Delta^\dagger \psi_R] + h.c. \nonumber \\
&  \quad = f_i [\overline{\psi^0_L} \phi^+_4 + \overline{\psi^{++}_L} \phi^{+++}_4 + \overline{\psi^+_L} \phi^{++}_4 + \overline{\psi^-_L} \phi^0_4 ] e_{R_i}  \\
& \quad + \frac{y_{D_i}}{3} [ \overline{e_{L_i}} (\sqrt3 \delta^{0*} \psi^-_R + 3 \delta^{--} \psi^+_R + \sqrt6 \delta^- \psi^0_R) 
+ \overline{\nu_{L_i}} (\sqrt3 \delta^{0*} \psi_R^0 + 3 \delta^{--} \psi^{++}_R + \sqrt6 \delta^- \psi^+_R) ]. \nonumber
\end{align}
Considering scalar mixing in Eqs.~\eqref{eq:scalar-mass-fields1}-\eqref{eq:scalar-mass-fields3} contributions to $\ell \to \ell' \gamma$ and muon $g-2$ are given by
one-loop diagram in Fig.~\ref{fig:diagram}.
Branching ratios (BRs) of LFV processes are written by  the following formula;
\begin{align}
 {\rm BR}(\ell_i\to\ell_j\gamma)= \frac{48\pi^3\alpha_{\rm em} C_{ij} }{(4\pi)^4{\rm G_F^2} m_{\ell_i}^2}\left(|a_{R_{ij}}|^2+|a_{L_{ij}}|^2\right).
 \end{align}
Dominant contributions to amplitudes $a_L$ and $a_R$ are given by
 \begin{align}
 a_{R_{ji}} &=-  y_{D_j} f_i (-1)^{k-1} \left[ \frac{\sqrt6 s_\alpha c_\alpha}3  D_a  F(m_{c_k},D_a)\right. \nn\\
&
\left.+  M_{\psi}
\left(s_\beta c_\beta\left(F(m_{d_k},M_{\psi}) - G(m_{d_k},M_{\psi})\right) + \frac{s_\gamma c_\gamma}{\sqrt{3}} G(m_{h_k},M_{\psi}) \right)\right] ,\\
 a_{L_{ji}} &=-  f^\dag_{j} y^\dag_{D_i} (-1)^{k-1} \left[  \frac{\sqrt6 s_\alpha c_\alpha}3  D_a  F(m_{c_k},D_a)\right. \nn\\
&
\left.+  M_{\psi}
\left(s_\beta c_\beta\left(F(m_{d_k},M_{\psi}) - G(m_{d_k},M_{\psi})\right) + \frac{s_\gamma c_\gamma}{\sqrt{3}} G(m_{h_k},M_{\psi}) \right)\right]  ,
\end{align}
where $a,k$ runs over $1,2$.
$m_{d_k},m_{c_k},m_{h_k}$ are respectively the mass eigenvalues for singly-charged bosons $H_{1,2}^{\pm}$ in Eq.~(17), doubly-charged ones $H_{1,2}^{\pm\pm}$ in Eq.~(18), and neutral ones $H_{1,2}^{0}$ in Eq.~(19),
and the loop functions are
\begin{align}
&F(m_a,m_b)\approx \frac{m_a^4 -m_b^4 + 2 m_a^2 m_b^2\ln\left(\frac{m_b^2}{m_a^2}\right) }{2(m_a^2 - m_b^2)^3},\\
&G(m_a,m_b)\approx -\frac{3m_a^4 +m_b^4 -4 m_a^2 m_b^2+2m_a^4\ln\left(\frac{m_b^2}{m_a^2}\right) }{2(m_a^2 - m_b^2)^3}.
\end{align}
The current experimental upper bounds on BRs of LFV processes are given 
by~\cite{MEG:2016leq,MEG:2013oxv}
  \begin{align}
  {\rm BR}(\mu\rightarrow e\gamma) &\leq4.2\times10^{-13},\quad 
  {\rm BR}(\tau\rightarrow \mu\gamma)\leq4.4\times10^{-8}, \quad  
  {\rm BR}(\tau\rightarrow e\gamma) \leq3.3\times10^{-8}~.
 \label{expLFV}
 \end{align}
 We impose these constraints in our numerical analysis below.
  Note also that there can be trilepton decay modes $\mu(\tau) \to \bar eee$ and $\tau \to \{\bar\mu \mu \mu, \bar\mu \mu e, \mu \mu\bar e, \bar\mu e e,\mu\bar e e \}$ mediated by doubly charged Higgs via Yukawa interaction of
 $\overline{L^c} \Delta L$. However corresponding Yukawa coupling is small in our case to realize neutrino mass since we choose $v_\Delta \sim 1$ GeV. Thus we can simply neglect these trilepton decays of $\mu$ and $\tau$.

Muon $g-2$; $\Delta a_\mu$, arises from the same diagram as LFVs and it is formulated by the following expression:
\begin{align}
&\Delta a_\mu \approx -\frac{m_\mu}{(4\pi)^2} [{a_{L_{22}}+a_{R_{22}}}] . \label{eq:G2-ZP}
\end{align}
The recent data tells us  $\Delta a_\mu= (24.9\pm4.9)\times 10^{-10}$~\cite{Muong-2:2021ojo,Muong-2:2023cdq}  at 1$\sigma$ C.L..
Note that $a_{L,R}$ does not have chiral suppression since vector-like lepton mass $M_\psi$ is picked up inside loop.
The simplest way to obtain the sizable muon $g-2$ is to set $f_{1,3}=y_{D_{1,3}}=0$, taking $f_2$ and $y_{D_2}$ to be order one. Then, we do not need to consider the constraints of LFVs. In the next subsection, we will show how it works through numerical analysis.

 Note also that there can be one-loop level vertex corrections for $Z \bar{\mu} \mu$ and $h \bar{\mu} \mu$ interactions associated with Yukawa coupling $y_{D_2}$ and $f_2$ realizing sizable muon $g-2$ that modify $Z(h) \to \bar\mu \mu$ decay modes. 
Typically we can satisfy experimental constraints when we have chiral enhancement for muon $g-2$. 
In fact we would have strong constraint from $Z \to \bar\mu \mu$ if we do not have chirality flip enhancement and it is one advantage of our model with such an enhancement.

\subsection{RG evolution of gauge couplings}
Here we briefly discuss renormalization group evolution of gauge coupling under existence of new particles. 
For illustration, we consider $U(1)_Y$ gauge coupling $g_Y$ and check in which scale it becomes strong coupling.
We find the energy evolution of $g_Y$, including contributions from $\{ \psi, H_4, \Delta \}$, such that 
\begin{align}
& \frac{1}{g^2_Y(\mu)} = \frac{1}{g^2_Y(m_{in})} - \frac{b_Y^{\rm SM}}{(4\pi)^2} \ln \left[\frac{\mu^2}{m_{in}^2} \right] - \theta(\mu - M) \frac{\Delta b_Y^{\psi} + \Delta b_Y^{H_4}   + \Delta b_Y^{\Delta}}{(4 \pi)^2} \ln \left[\frac{\mu^2}{M^2} \right], \\
& \Delta b_Y^{\psi}  = \frac{1}{10}, \ \Delta b_Y^{H_4} = \frac{9}{20}, \  \Delta b_Y^{\Delta} = \frac{3}{20},
\end{align}
where $\mu$ is a reference energy scale, and we choose $m_{in} = m_Z$ and $M = 1$ TeV; $m_{in}$ and $M$ are initial and threshold mass scale respectively.
As a result, we find that $g_Y$ becomes $\mathcal{O}(1)$ around $\mu = 10^{32}$ GeV which is much larger than Planck scale. 
The evolution of $SU(2)_L$ gauge coupling does not change significantly and electroweak gauge couplings do not become strong interaction below Planck scale in the model.

\section{Numerical analysis and phenomenology}
\begin{figure}[tb]\begin{center}
\includegraphics[width=10cm]{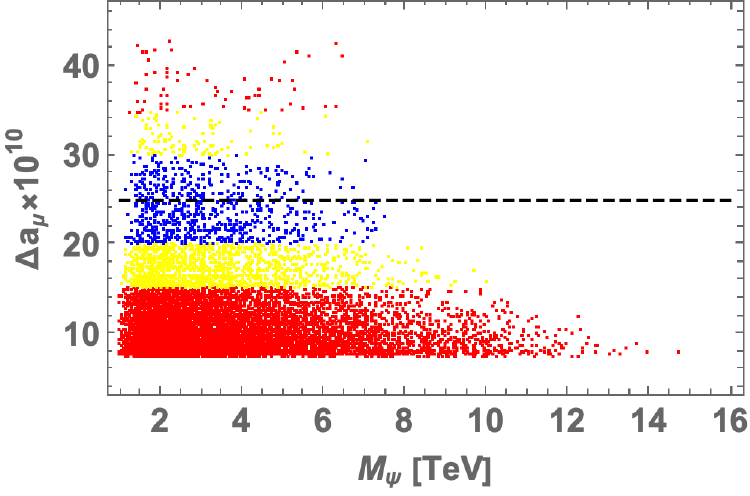}
\caption{Random plot of muon $g-2$ in terms of mass parameter $M_\psi$ within the range of Eq.~(\ref{eq:pr}) where we imposed LFV constraints.
The black dashed line represents the best fit value of muon $g-2$, the blue points are within 1$\sigma$, the yellow ones are within 2$\sigma$, and red ones are within 3$\sigma$ of experimental value.
}   \label{fig:mamu}
\end{center}\end{figure}
\begin{figure}[tb]\begin{center}
\includegraphics[width=8cm]{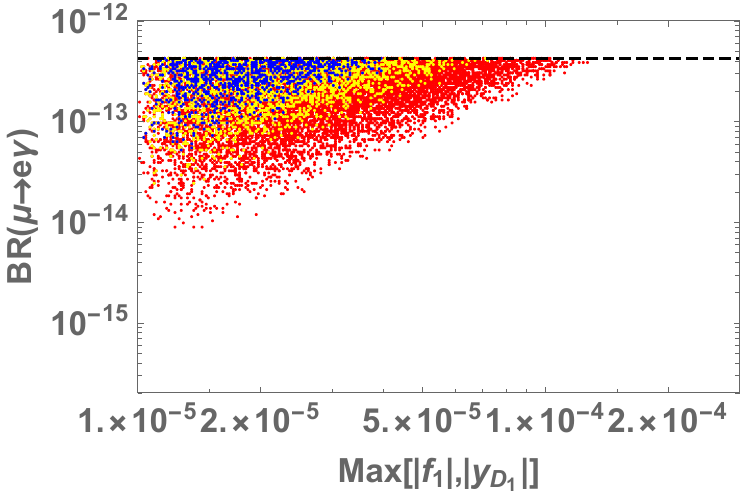} \
\includegraphics[width=8cm]{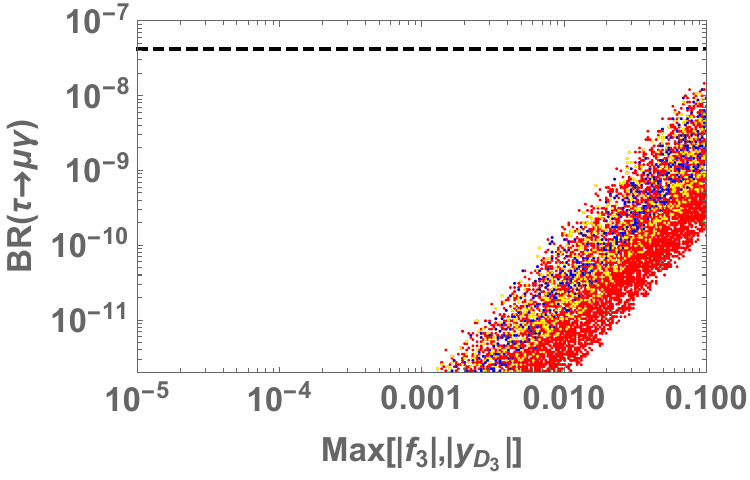}
\caption{Left and right plots show $BR(\mu \to e \gamma)$ and $BR(\tau \to \mu \gamma)$ as functions of Max$[|f_1|, |y_{D_1}|]$ and Max$[|f_3|, |y_{D_3}|]$ 
for parameter sets satisfying muon $g-2$ within $3 \sigma$; the color of points is the same as in Fig.~\ref{fig:mamu}. The horizontal dashed line indicates current upper bound of the BRs.
}   \label{fig:LFV}
\end{center}\end{figure}

In this section we carry out numerical analysis by scanning free parameters and explore the region to explain muon $g-2$ taking into account LFV constraints.
Then we consider collider physics focusing on production of multiply-charged fermions and scalar bosons.

\subsection{Numerical analyses on muon $g-2$}
Now that the formalulations have been done, we carry out numerical analysis taking into account LFV constraints and muon $g-2$.
At first, we randomly select the following input parameters:
\begin{align}
& M_\psi  \supset [10^3,10^5] \ {\rm GeV},\quad  m  \supset [0.01,10] \ {\rm GeV},\quad \mu_{H_4} = 1.2 M_\psi, \quad \mu_\Delta = 0.8 M_\psi,\nn\\  
&  \mu_1  \supset [100,\mu_{H_4}] \ {\rm GeV}, \quad \{ |f_{2}|, |y_{D_{2}}| \} \supset [0.1,2.0],\quad \{ |f_{1,3}|, |y_{D_{1,3}}| \} \supset [10^{-5},0.1],
\label{eq:pr}
 \end{align}  
 where we chose $|f_2|$ and $|y_{D_2}|$ to be larger than other Yukawa couplings so that we have sizable muon $g-2$.
Note that splittings of masses of components in the same scalar multiplets are small and we can evade the constraints from oblique parameters~\cite{Peskin:1990zt}. 
 Also we take $\mu_2 =0$ for simplicity.
 
Fig.~\ref{fig:mamu} represents the values of muon $g-2$ in terms of mass parameter $M_\psi$ where each point corresponds to one parameter sets within the range of Eq.~(\ref{eq:pr}) allowed by LFV constraints that satisfy a value of muon $g-2$ in $3 \sigma$.
The black dashed line shows the best fit value of muon $g-2$, the blue points are within 1$\sigma$, the yellow ones are within 2$\sigma$, and red ones are within 3$\sigma$ of experimental value.
 We thus find that $M_\psi \lesssim 15(8.5)$ TeV is preferred to obtain muon $g-2$ within $3(1) \sigma$ C.L. in our scenario.
In addition we show branching ratios of LFV processes for the same parameter sets in Fig.~\ref{fig:LFV} where the left and right plots represent $BR(\mu \to e \gamma)$ and $BR(\tau \to \mu \gamma)$ as functions of Max$[|f_1|, |y_{D_1}|]$ and Max$[|f_3|, |y_{D_3}|]$, and the color of points is the same as Fig.~\ref{fig:mamu}. 
It is found that $|f_{1}|(|y_{D_1}|)$ should be smaller than $\sim 10^{-4}$ to avoid stringent constraint from $BR(\mu \to e \gamma)$ while constraint on $f_3 (y_{D_3})$ is much looser.
Here we omitted a plot for $BR(\tau \to e \gamma)$ since it is not correlated to muon $g-2$ and it tends to be much smaller than experimental limit.

\begin{figure}[tb]\begin{center}
\includegraphics[width=5.25cm]{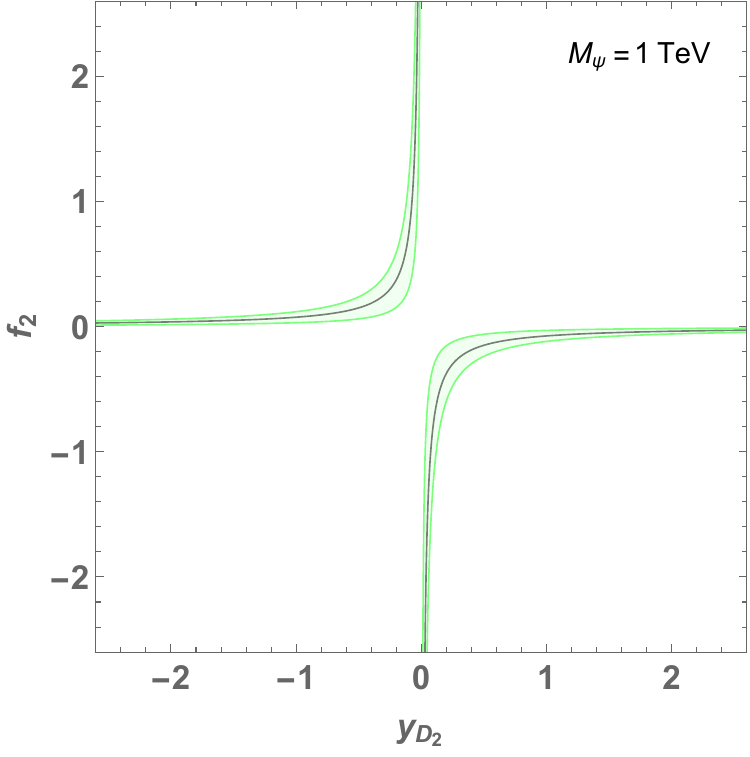} \
\includegraphics[width=5.25cm]{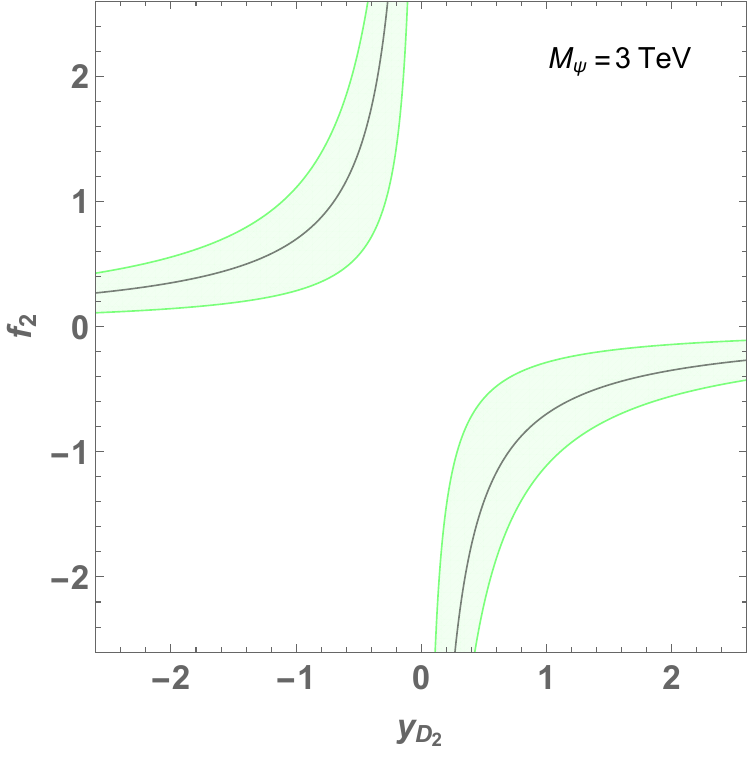} \
\includegraphics[width=5.25cm]{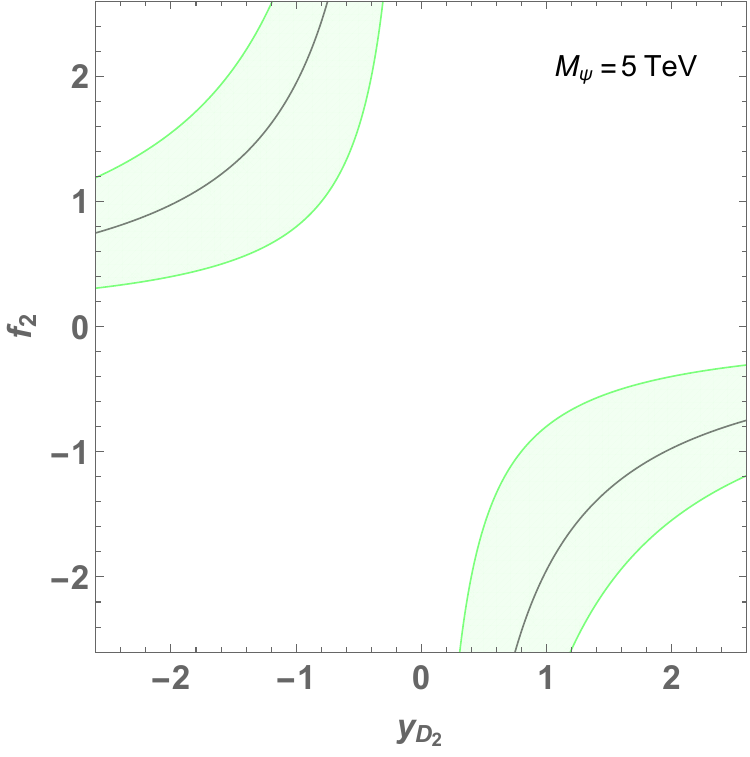}
\caption{
The region realizing muon $g-2$ within 3$\sigma$ C.L. for $M_\psi = \{1, 3, 5\}$ TeV on $\{y_{D_2}, f_2 \}$ plane where we chose $\mu_1=\mu_{H_4} = 1.2 M_\psi$, $\mu_\Delta = 0.8 M_\psi$, $m = 10$ GeV and $y_{D_{1,3}} = f_{1,3} =0$. 
The parameters on the black curve provide best fit value of the muon $g-2$.
}   \label{fig:f2amu}
\end{center}\end{figure}
Next, we also demonstrate a simple realization to get sizable muon $g-2$ setting $f_{1,3}=y_{D_{1,3}}=0$, taking $f_2$ and $y_{D_2}$ to be free parameters. This is because we can enhance the muon $g-2$ without inducing LFVs.
Fig.~\ref{fig:f2amu} represents  region realizing muon $g-2$ within $3 \sigma$ C.L. on the parameter space of the valid Yukawa coupling $y_{D_2}$ and $f_2$  fixing the other input parameters as follows; $M_\psi =\{1, 3, 5\}$ TeV as indicated on the plots, $\mu_{H_4} = 1.2 M_\psi$, $\mu_\Delta = 0.8 M_\psi$, $m=10$ GeV, and $\mu_1 = \mu_{H_4}$.
The black solid curves represent the parameter region providing the best fit value of muon $g-2$. 
One finds that less than order one Yukawa couplings are enough to find the best fit value of muon $g-2$ even when the fermion mass is of the order 3 TeV.

\subsection{ Collider physics}
\begin{figure}[tb]
\begin{center}
\includegraphics[width=5.25cm]{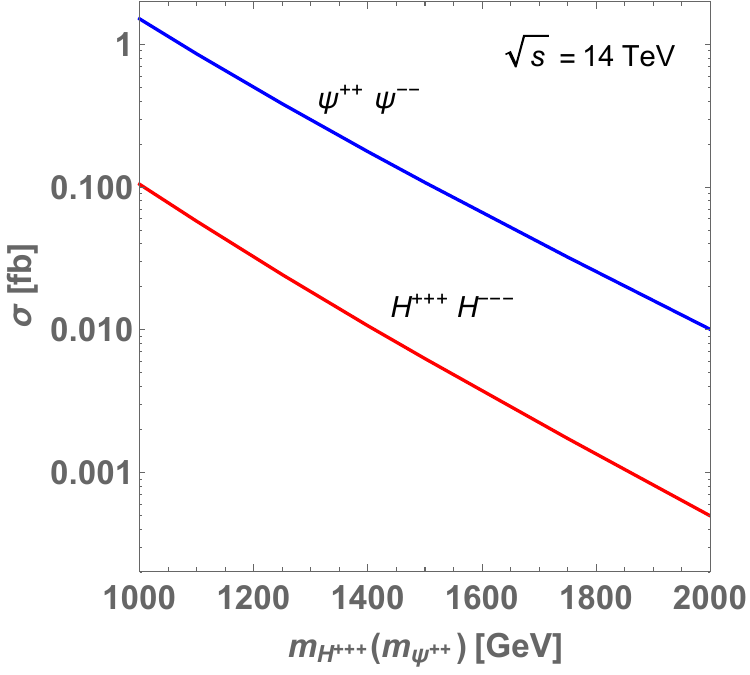} \
\includegraphics[width=5.25cm]{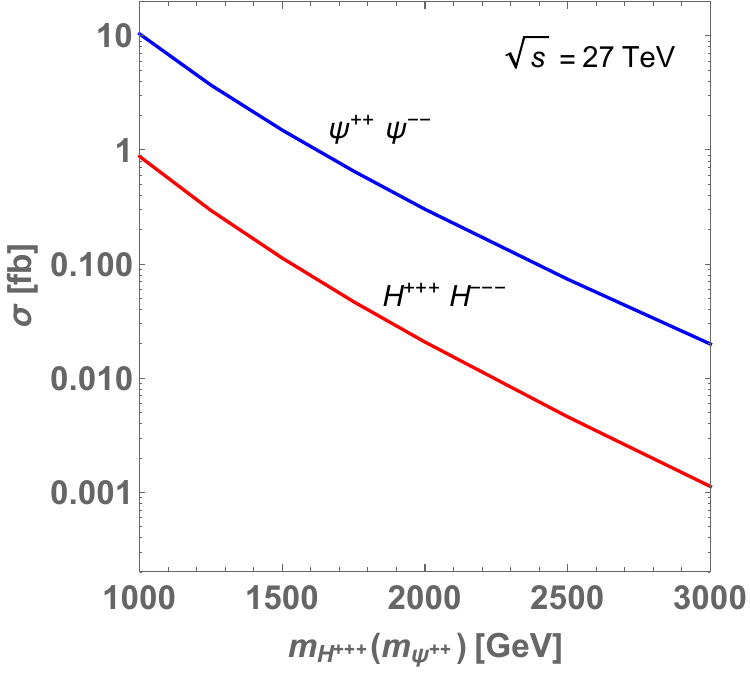} \
\includegraphics[width=5.25cm]{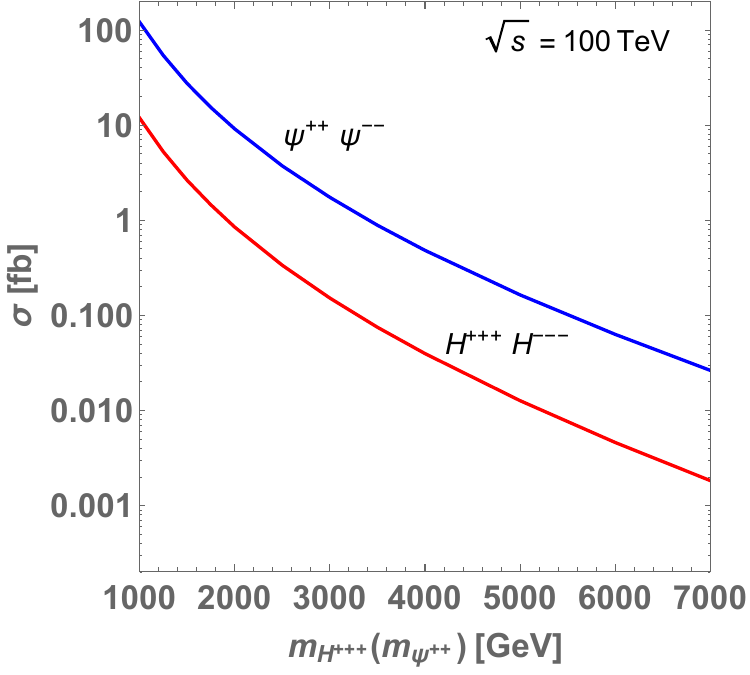}
\caption{Cross sections for $pp \to H^{+++}H^{---}$ and $pp \to \psi^{++} \psi^{--}$ with center of mass energy 14 TeV, 27 TeV and 100 TeV.}
\label{fig:CX}
\end{center}\end{figure}

Here we briefly discuss collider signature of the model focusing on the pair productions of new particles with the highest electric charge in $H_4$ and $\psi$. 
They can be produced via electroweak gauge interactions that are given by
\begin{align}
& (D_\mu H_4)^\dagger (D^\mu H_4) \supset  i \left[ \frac{g}{c_W} \left( \frac32 - 3 s^2_W \right) Z_\mu + 3 e A_\mu \right] (\partial^\mu H^{+++} H^{---} - \partial^\mu H^{---}  H^{+++}) \\
& \bar{\psi} i \gamma^\mu D_\mu \psi \supset \overline{\psi^{++}} \gamma^\mu \left[ \frac{g}{c_W} \left( \frac32 - 2 s^2_W \right)  Z_\mu + 2 e A_\mu \right] \psi^{++},
\end{align}
where we omitted other terms which are irrelevant in our calculation below.
We consider the production processes
\begin{align}
& pp \to Z/\gamma \to H^{+++} H^{---}, \\
& pp \to Z/\gamma \to \overline{\psi^{++}} \psi^{++},
\end{align}
in a hadron collider experiment.
Here we estimate the production cross sections in use of {\it CalcHEP 3.8}~\cite{Belyaev:2012qa} package implementing the relevant interactions applying the CTEQ6 parton distribution functions (PDFs)~\cite{Nadolsky:2008zw}.
In Fig.~\ref{fig:CX}, the cross sections are shown as functions of exotic charged particle masses for center of mass energy $\sqrt{s} = 14, 27$ and 100 TeV as reference values.
We find that $\overline{\psi^{++}} \psi^{++}$ production cross section is larger than that of $H^{+++}H^{---}$ by one order.

Exotic charged particles can decay via Yukawa couplings in Eq.~\eqref{Eq:yuk}.
Here we assume relation among mass parameters as $\mu_\Delta < M_\psi < \mu_{H_4}$ for illustration.
Then dominant decay modes of $H^{+++}$ and $\psi^{++}$ are respectively $H^{+++} \to \ell^+ \psi^{++}$ and $\psi^{++} \to \nu_\ell \delta^{++}$ 
where the decay widths are given by
\begin{align}
 \Gamma(H^{+++} \to \ell^+ \psi^{++}) &= \frac{f_\ell^2}{16 \pi} m_{H^{+++}} \left( 1 - \frac{M_\psi^2}{m^2_{H^{+++}}} \right)^2, \\
 \Gamma(\psi^{++} \to \nu_\ell \delta^{++}) &=   \frac{y_\ell^2}{32 \pi } M_\psi \left(1 - \frac{m^2_{\delta^{++}}}{M^2_\psi} \right)^2. 
\end{align}
Note that here we consider $\delta^{++} \simeq H_1^{++}$ assuming small mixing angle for simplicity.
 The mixing angle $\beta$ between doubly charged scalar is roughly estimated by $|\beta| \simeq \Delta M^2_{++}/\mu^2_{H_4} \sim \mu_1 v_h/\mu^2_{H_4}$, 
and it can be small; $|\beta | < 0.1$ is realized $\mu_1 \lesssim \mu_{H_4}/2$ for $\mu_{H_4} = 1.2$ TeV and the approximation above is fine. 
Note that the branching ratio of these processes are $1$ when the mass difference between components in the multiplets are small enough.
In addition $\delta^{++}$ dominantly decays into $W^+ W^+$ mode considering $v_\Delta \sim \mathcal{O}(1)$ GeV. 
Thus decay chains provide signature from $H^{+++}H^{---}$ and $\psi^{++} \psi^{--}$ such that 
\begin{align}
& \psi^{++} \overline{\psi^{++}} \to \nu \bar{\nu} \delta^{++} \delta^{--} \to \nu \bar{\nu} W^+ W^+ W^- W^-, \\
& H^{+++}H^{---} \to \ell^+ \ell^- \psi^{++} \overline{\psi^{++}} \to \ell^+ \ell^- \nu \bar{\nu} \delta^{++} \delta^{--} \to \ell^+ \ell^- \nu \bar{\nu} W^+ W^+ W^- W^-, 
\end{align}
where $\nu (\bar \nu)$ indicates any flavor of neutrino (anti-neutrino).
For signals, we consider one pair of same sign $W$ bosons decays into leptons while the other same sign pair decays into jets.
The signals at detectors are 
\begin{align}
\label{eq:signal1}
& \text{Signal 1:} \quad \ell^\pm \ell^\pm 4 j \slashed{E}_T, \\
\label{eq:signal2}
& \text{Signal 2:} \quad \ell^\pm \ell^\pm \ell^\pm \ell^\mp 4 j \slashed{E}_T,
\end{align}
where $j$ and $\slashed{E}_T$ indicate jet and missing transverse momentum.
In Table~\ref{tab:event}, we provide expected number of events given by the products of luminosity $L$, production cross section $\sigma$ and BRs for corresponding final state, 
for some benchmark points (BPs) of $M_\psi (m_{H^{+++}})$ assuming integrated luminosity of $L = 1$ ab$^{-1}$ and $m_{\delta^{++}} < M_\psi$ to allow the decay mode of $\psi^{++} \to \delta^{++} \nu$. 
We thus find that mass scale of around 1 TeV can be explored by $\sqrt{s} = 14$ TeV with sufficient integrated luminosity that will be achieved by High-Luminosity-LHC experiment.
In particular, the signal from $\overline{\psi^{++}} \psi^{++}$ is promising since the cross section is larger than that of $H^{+++} H^{---}$.

\begin{table}[t!]
\begin{tabular}{|c||c|c|c|c||c|c|c|c||c|c|c|c|}\hline
$\sqrt{s}$ [TeV] & \multicolumn{4}{|c||}{14} &  \multicolumn{4}{|c||}{27} &  \multicolumn{4}{|c|}{100} \\ \hline
$M_\psi (m_{H^{+++}})$ [GeV] & 1000 & 1250 & 1500 & 1750 & 1000 & 1500 & 2000 & 2500 & 2000 & 3000 & 4000 & 6000 \\ \hline
$L \cdot \sigma \cdot {\rm BR}$ [Signal 1] & 324. & 82. & 23. & 7. & 2221. & 319. & 65. & 16. & 1954. & 373. & 103. & 14. \\ 
$L \cdot \sigma \cdot {\rm BR}$ [Signal 2] & 22. & 5. & 1. & 0. & 188. & 24. & 4. & 1. & 181. & 33. & 8. & 1. \\ \hline
\end{tabular}
\caption{Expected number of events for our signals in some BPs of $M_\psi (m_{H^{+++}}$)  with integrated luminosity of $L = 1$ ab$^{-1}$. }\label{tab:event}
\end{table}

{
Next we carry out simple numerical simulation study to examine possibility of testing our signal at the LHC 14 TeV including detector effect.
For illustration, the signal 1 in Eq.~\eqref{eq:signal1} is considered since the expected number of events is larger than the other signal.
We also fix doubly charged scalar mass as $m_\delta \equiv m_{\delta^{\pm \pm}} = 500$ GeV and $H^{\pm \pm \pm}$ mass as 2000 GeV for simplicity.
The possible SM background(BG) processes for the signal are as follows: 
\begin{align}
 pp \to & \large\{ ZZZ, \ ZW^+W^-, \ ZZW^\pm, \ ZZZZ, \ W^+W^-W^+W^-,  \nonumber \\
& \ ZZW^+W^-, \ ZZZW^\pm, \ W^\pm W^\pm q q \large\},  \label{eq:BGs}
\end{align} 
where $q$ indicates any quarks and these processes provides charged leptons, missing transverse momentum and jets after the decay of gauge bosons.  Note that it is more straightforward and even precise to directly generate charged leptons, jets and missing transverse energy events as BG in simulation study. However it is difficult to generate final states containing many particles such as 8 particle states $\ell^\pm \ell^\pm 4j \nu \nu$. We thus generate final states in Eq.~\eqref{eq:BGs} in our simulation study; note that in that case the number of BG events would be underestimated bit.
We estimate the cross sections of these BG processes and find 
\begin{align}
& \sigma(pp \to ZZZ) = 10.3 \ {\rm fb}, \  \sigma(pp \to ZW^+W^-) = 9.44 \ {\rm fb}, \ \sigma(pp \to ZZW^+) = 19.9 \ {\rm fb} \nonumber \\
& \sigma(pp \to ZZW^-) = 10.4 \ {\rm fb}, \ \sigma(pp \to ZZZZ) = 1.95 \times 10^{-2} \ {\rm fb}, \nonumber \\
& \sigma(pp \to W^+W^-W^+W^-) = 0.571 \ {\rm fb}, \ \sigma(pp \to ZZW^+W^-) = 0.436 \ {\rm fb}, \nonumber \\
& \sigma(pp \to ZZZW^+) = 4.21 \times 10^{-2} \ {\rm fb}, \ \sigma(pp \to ZZZW^-) = 1.87 \times 10^{-2} \ {\rm fb}, \nonumber \\
&  \sigma(pp \to W^+W^+ q q) = 215 \ {\rm fb}, \ \sigma(pp \to W^- W^- q q) = 93.4 \ {\rm fb},
\end{align}
that are estimated by {\tt MADGRAPH5}~\cite{Alwall:2014hca}.
Then we perform numerical simulation generating events for signal and BG using {\tt MADGRAPH5} by implementing the model in use of FeynRules 2.0 \cite{Alloul:2013bka}.
The events are passed to {\tt PYTHIA\,8}~\cite{Sjostrand:2014zea} to deal with hadronization effects,  the  initial-state radiation (ISR) and final-state radiation (FSR) effects and the decays of SM particles, 
and we apply {\tt Delphes}~\cite{deFavereau:2013fsa} for detector level simulation.
We apply selection of events at the detector level such that
\begin{align}
\label{eq:selection}
\ell^\pm \ell^\pm +  \text{at least 3 jets}.
\end{align} 
Here we consider minimum number of jets is three not to much reduce signal events.

\begin{figure}[tb]
\begin{center}
\includegraphics[width=5.25cm]{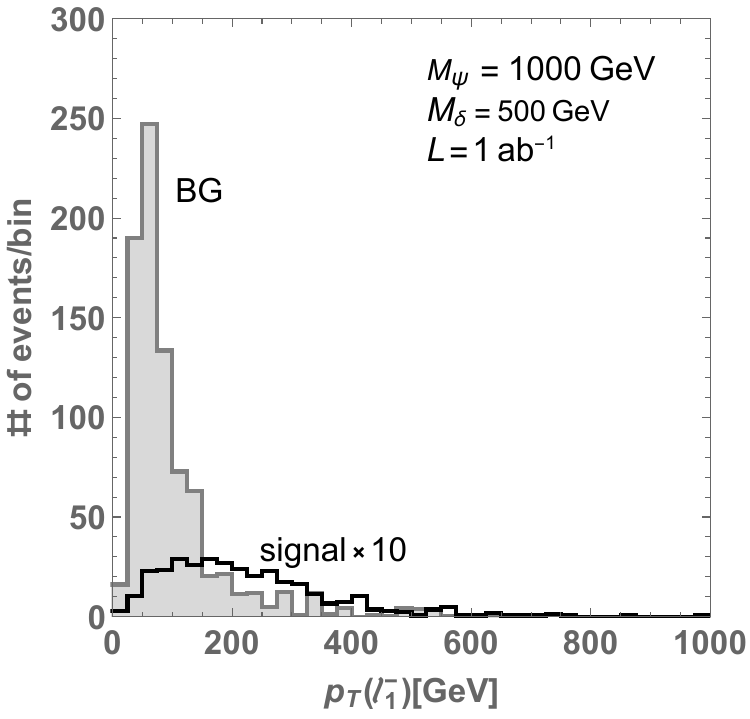} \
\includegraphics[width=5.25cm]{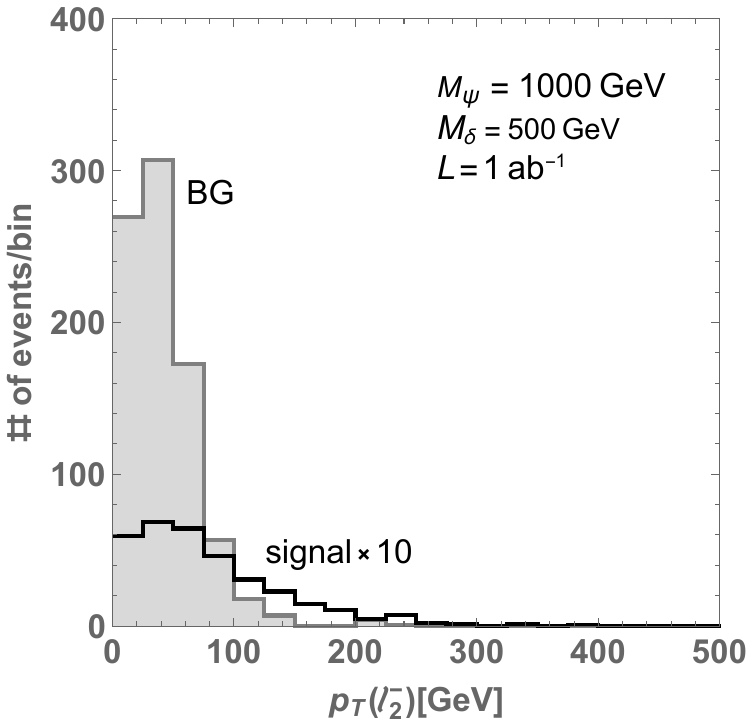} \
\includegraphics[width=5.25cm]{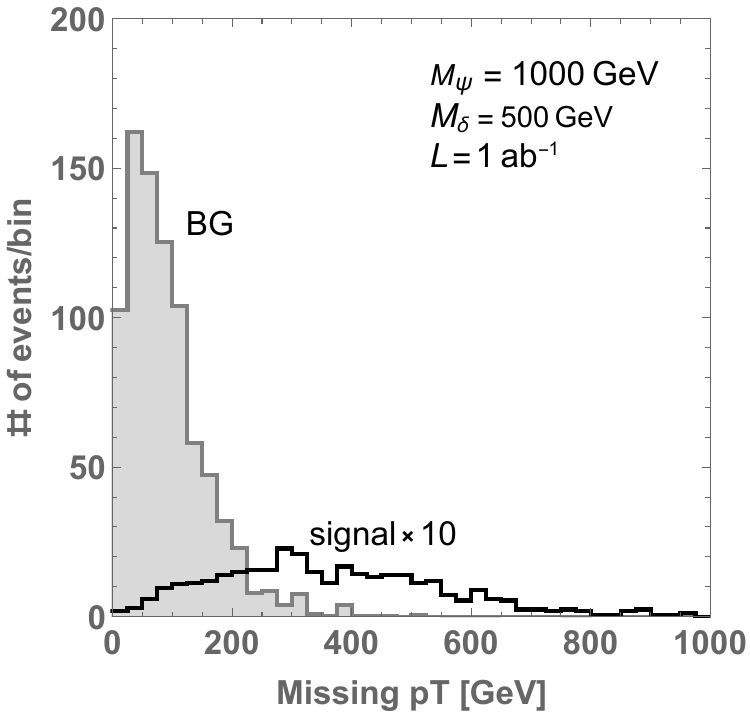}
\caption{Left: Distribution of transverse momentum of $\ell^-_1$. Center: Distribution of transverse momentum of $\ell^-_2$. Right: Distribution of missing transverse momentum. The solid black histogram indicates signal while the gray filled histogram is for BG events. Here $\ell_{1(2)}$ is charged lepton with the (second) highest transverse momentum. The applied luminosity and masses of new particles are given in the plots. }
\label{fig:distribution1}
\end{center}\end{figure}

\begin{figure}[tb]
\begin{center}
\includegraphics[width=5.25cm]{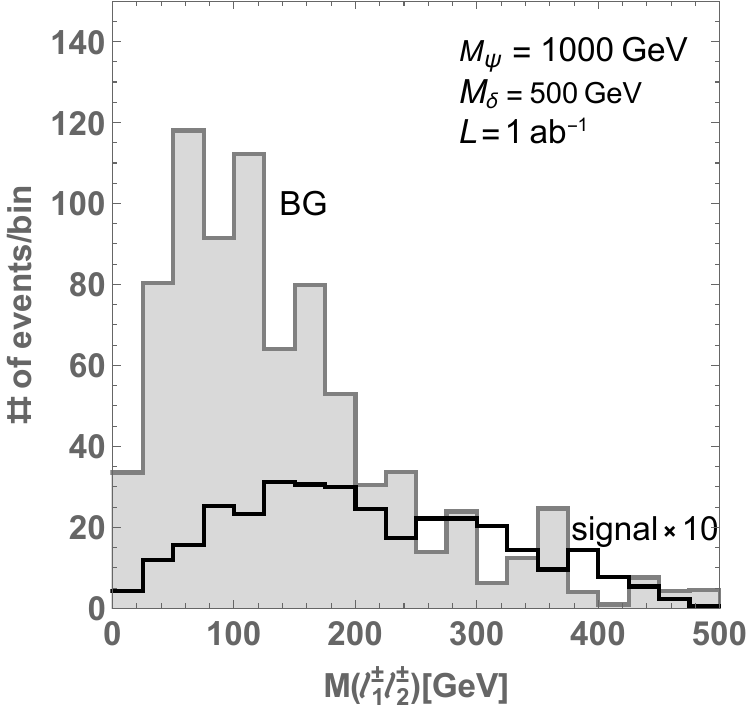} \
\includegraphics[width=5.25cm]{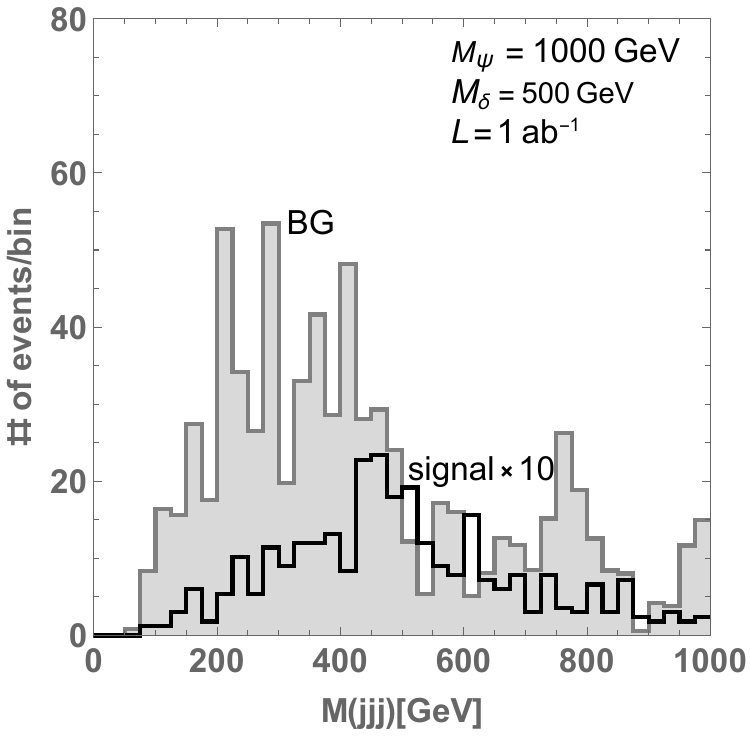} 
\caption{Left: Distribution of invariant mass of same sign charged lepton.  Right: Distribution of invariant mass of three jets. The indication of histograms is the same as Fig.~\ref{fig:distribution1}. }
\label{fig:distribution2}
\end{center}\end{figure}

\begin{center} 
\begin{table}[t]
\begin{tabular}{|c||c|c|c|c|c|c|c|c|c|c|c|c|}\hline
  & $N_{\rm signal}$ & $N_{ZZZ}$ & $N_{ZW^+W^-}$ & $N_{ZZW^\pm}$  & $N_{4Z}$ & $N_{2W^+W^-}$ & $N_{ZZW^+W^-}$ & $N_{ZZZW^\pm}$ & $N_{W^\pm W^\pm qq}$ & $S$   \\ \hline 
Before cuts &  33.4 & $7.83$ & $13.6$  & $53.5$ & $0.0819$ & $10.3$ & $3.02$ & $0.313$ & $747$ & $1.15$ \\ \hline
With cuts  & 15.4 & 0.001  & $0.378$ & $0.832$ & $0.00234$ & $0.458$ & $0.192$ & $0.00524$  & $3.74$ & $5.67$   \\ \hline
\end{tabular}
\caption{Number of events before and after kinematical cuts for event selection of $\ell^- \ell^- +$jets. }
\label{tab:Event}
\end{table}
\end{center}

In Fig.~\ref{fig:distribution1}, we show kinetic distributions for signal and BG events where we apply integrated luminosity $L = 1$ab$^{-1}$ and consider $\ell^- \ell^-$ case in event section Eq.~\eqref{eq:selection} (results for $\ell^+ \ell^+$ case are the almost same); 
left plot is distribution of transverse momentum of $\ell^-_1$ ($\ell_{1(2)}$ is charged lepton with the (second) highest transverse momentum), center plot is distribution of transverse momentum of $\ell^-_2$, right plot is distribution of missing transverse momentum. The solid black histogram indicates signal while the gray filled histogram is for BG events where all BG events are summed up. 
Here number of events are estimated as $N_{\rm event} = L \sigma N_{\rm Selected}/N_{\rm Generated}$ 
where $N_{\rm Select}$ is number of events after selection, $N_{\rm Generated}$ is number of originally generated events by {\tt MADEVENT5}, $\sigma$ is a cross section for each process and $L$ is integrated luminosity.
We find that $P_T(\ell^-)$ and missing transverse momentum of signal tend to be larger than those of BG.
We also show distributions for invariant mass of the same sign leptons and three jets in Fig.~\ref{fig:distribution2} where we consider three jets since we chose minimal number of jets as three. 
It is found that we the signal distribution of three jet invariant mass is centered around the mass of $\delta^{\pm \pm}$ since jets in signal events mainly come from the decay chain $\delta^{\pm \pm} \to W^\pm W^\pm \to jjjj$.

\begin{figure}[tb]
\begin{center}
\includegraphics[width=8.25cm]{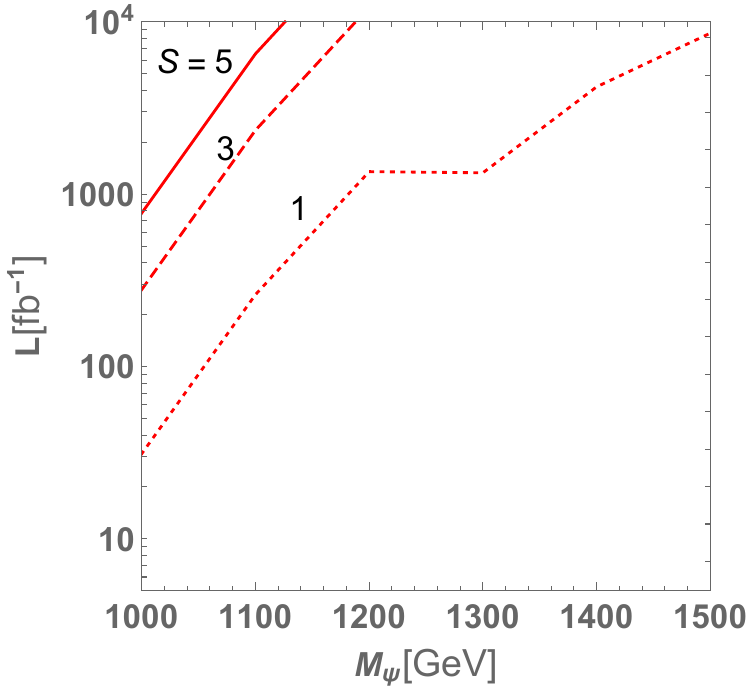} \
\caption{The integrated luminosity to realize discovery significance $S=1,3,5$ where kinematical cuts Eq.~\eqref{eq:selection} are applied and both $\ell^+ \ell^+$ and $\ell^- \ell^-$ cases are summed up.  The center of mass energy is $\sqrt{s} = 14$ TeV.}
\label{fig:significance}
\end{center}\end{figure}

Based on the kinetic distributions, we apply kinematical cuts 
\begin{equation}
 p_T(\ell^\pm_1) > 150 \ {\rm GeV}, \ \slashed{E}_T > 250 \ {\rm GeV},  
\label{eq:cut}
\end{equation}
where $\slashed{E}_T$ is the missing transverse energy(momentum)  that is the same as missing $p_T$ in Fig.~\ref{fig:distribution1}.
For illustration, we show the number of signal events before and after the cut Eq.~\eqref{eq:cut} for $\ell^- \ell^-$ case and estimated the discovery significance in use of the formula 
\begin{equation}
S = \sqrt{2 \left[ (N_S + N_{BG}  ) \ln \left( 1 + \frac{N_S}{N_{BG}} \right) - N_S\right]},
\end{equation}
where where $N_S$ and $N_{BG}$ are respectively number of signal and total BG events.
We find that the kinematical cuts can reduce number of backgrounds significantly while the number of signal events is not much reduced.
 In particular, we find the cut of missing transverse energy improves signal to background ratio effectively.
 Then we can get larger significance $S$ after imposing cuts as show in Table~\ref{tab:Event}.
Finally, in Fig.~\ref{fig:significance}, we show required integrated luminosity to realize discovery significance $S=1,3,5$ where kinematical cuts  Eq.~\eqref{eq:cut} are applied and both $\ell^+ \ell^+$ and $\ell^- \ell^-$ cases are summed up.
It is found that $M_\psi \lesssim 1060$ GeV region can be in reach of discovery at the high-luminosity LHC with $L = 3000$fb$^{-1}$.
We expect more mass region explaining muon $g-2$ could be tested if we realize a higher energy experiment like 100 TeV collider in future~\cite{FCC:2018byv}.

}

\section{Summary and discussion}

In this paper we have proposed a simple extension of the SM without additional symmetry by introducing large $SU(2)_L$ multiplet fields such as 
quartet vector-like fermion as well as quartet and triplet scalar fields.
These multiplet fields can induce sizable muon $g-2$ due to new Yukawa couplings at one-loop level where we do not have chiral suppression by light lepton mass as we pick up heavy fermion mass term changing chirality inside a loop diagram.
The triplet scalar field can induce neutrino masses after developing its VEV  by type-II seesaw mechanism.

We have carried out numerical analysis searching for parameter space that explains muon $g-2$ and allowed by LFV constraints.
It has been then found that muon $g-2$ can be explained when the new multiplets mass scale are less than around $\mathcal{O}(10)$ TeV.
We also discussed collider physics focusing on production of multiply-charged fermions/scalars via proton-proton collision.
 We have carried out simulation study for $\sqrt{s} = 14$ TeV fixing some parameters for illustration. 
The signal/background events are generated and we find relevant kinematical cuts to reduce background via kinematical distributions. 
Then we have investigated the effect of cuts and have shown discovery significance after imposing the cuts. 
Mass scale of 1 TeV can be explored by $\sqrt{s} = 14$ TeV with sufficient integrated luminosity that will be achieved by High-Luminosity-LHC experiment.
Also most of mass region explaining muon $g-2$ could be tested if we realize a higher energy experiment like 100 TeV collider.

\section*{Acknowledgments}
This research was supported by an appointment to the JRG Program at the APCTP through the Science and Technology Promotion Fund and Lottery Fund of the Korean Government. This was also supported by the Korean Local Governments - Gyeongsangbuk-do Province and Pohang City (H.O.). 
The work was also supported by the Fundamental Research Funds for the Central Universities (T.~N.).


\begin{thebibliography}{99}


  



\bibitem{Muong-2:2021ojo}
B.~Abi \textit{et al.} [Muon g-2],
Phys. Rev. Lett. \textbf{126} (2021) no.14, 141801
doi:10.1103/PhysRevLett.126.141801
[arXiv:2104.03281 [hep-ex]].

\bibitem{Muong-2:2023cdq}
D.~P.~Aguillard \textit{et al.} [Muon g-2],
Phys. Rev. Lett. \textbf{131} (2023) no.16, 161802
[arXiv:2308.06230 [hep-ex]].


\bibitem{Hagiwara:2011af} 
  K.~Hagiwara, R.~Liao, A.~D.~Martin, D.~Nomura and T.~Teubner,
  J.\ Phys.\ G {\bf 38}, 085003 (2011)
  [arXiv:1105.3149 [hep-ph]].
  

\bibitem{Aoyama:2012wk}
T.~Aoyama, M.~Hayakawa, T.~Kinoshita and M.~Nio,
Phys. Rev. Lett. \textbf{109}, 111808 (2012)
doi:10.1103/PhysRevLett.109.111808
[arXiv:1205.5370 [hep-ph]].

\bibitem{Aoyama:2019ryr}
T.~Aoyama, T.~Kinoshita and M.~Nio,
Atoms \textbf{7}, no.1, 28 (2019)
doi:10.3390/atoms7010028

\bibitem{Czarnecki:2002nt}
A.~Czarnecki, W.~J.~Marciano and A.~Vainshtein,
Phys. Rev. D \textbf{67}, 073006 (2003)
[erratum: Phys. Rev. D \textbf{73}, 119901 (2006)]
doi:10.1103/PhysRevD.67.073006
[arXiv:hep-ph/0212229 [hep-ph]].

\bibitem{Gnendiger:2013pva}
C.~Gnendiger, D.~St\"ockinger and H.~St\"ockinger-Kim,
Phys. Rev. D \textbf{88}, 053005 (2013)
doi:10.1103/PhysRevD.88.053005
[arXiv:1306.5546 [hep-ph]].

\bibitem{Keshavarzi:2018mgv}
A.~Keshavarzi, D.~Nomura and T.~Teubner,
Phys. Rev. D \textbf{97}, no.11, 114025 (2018)
doi:10.1103/PhysRevD.97.114025
[arXiv:1802.02995 [hep-ph]].

\bibitem{Colangelo:2018mtw}
G.~Colangelo, M.~Hoferichter and P.~Stoffer,
JHEP \textbf{02}, 006 (2019)
doi:10.1007/JHEP02(2019)006
[arXiv:1810.00007 [hep-ph]].

\bibitem{Hoferichter:2019mqg}
M.~Hoferichter, B.~L.~Hoid and B.~Kubis,
JHEP \textbf{08}, 137 (2019)
doi:10.1007/JHEP08(2019)137
[arXiv:1907.01556 [hep-ph]].

\bibitem{Keshavarzi:2019abf}
A.~Keshavarzi, D.~Nomura and T.~Teubner,
Phys. Rev. D \textbf{101}, no.1, 014029 (2020)
doi:10.1103/PhysRevD.101.014029
[arXiv:1911.00367 [hep-ph]].

\bibitem{Kurz:2014wya}
A.~Kurz, T.~Liu, P.~Marquard and M.~Steinhauser,
Phys. Lett. B \textbf{734}, 144-147 (2014)
doi:10.1016/j.physletb.2014.05.043
[arXiv:1403.6400 [hep-ph]].

\bibitem{Melnikov:2003xd}
K.~Melnikov and A.~Vainshtein,
Phys. Rev. D \textbf{70}, 113006 (2004)
doi:10.1103/PhysRevD.70.113006
[arXiv:hep-ph/0312226 [hep-ph]].

\bibitem{Masjuan:2017tvw}
P.~Masjuan and P.~Sanchez-Puertas,
Phys. Rev. D \textbf{95}, no.5, 054026 (2017)
doi:10.1103/PhysRevD.95.054026
[arXiv:1701.05829 [hep-ph]].

\bibitem{Colangelo:2017fiz}
G.~Colangelo, M.~Hoferichter, M.~Procura and P.~Stoffer,
JHEP \textbf{04}, 161 (2017)
doi:10.1007/JHEP04(2017)161
[arXiv:1702.07347 [hep-ph]].

\bibitem{Hoferichter:2018kwz}
M.~Hoferichter, B.~L.~Hoid, B.~Kubis, S.~Leupold and S.~P.~Schneider,
JHEP \textbf{10}, 141 (2018)
doi:10.1007/JHEP10(2018)141
[arXiv:1808.04823 [hep-ph]].

\bibitem{Gerardin:2019vio}
A.~G\'erardin, H.~B.~Meyer and A.~Nyffeler,
Phys. Rev. D \textbf{100}, no.3, 034520 (2019)
doi:10.1103/PhysRevD.100.034520
[arXiv:1903.09471 [hep-lat]].

\bibitem{Bijnens:2019ghy}
J.~Bijnens, N.~Hermansson-Truedsson and A.~Rodr\'\i{}guez-S\'anchez,
Phys. Lett. B \textbf{798}, 134994 (2019)
doi:10.1016/j.physletb.2019.134994
[arXiv:1908.03331 [hep-ph]].

\bibitem{Colangelo:2019uex}
G.~Colangelo, F.~Hagelstein, M.~Hoferichter, L.~Laub and P.~Stoffer,
JHEP \textbf{03}, 101 (2020)
doi:10.1007/JHEP03(2020)101
[arXiv:1910.13432 [hep-ph]].

\bibitem{Blum:2019ugy}
T.~Blum, N.~Christ, M.~Hayakawa, T.~Izubuchi, L.~Jin, C.~Jung and C.~Lehner,
Phys. Rev. Lett. \textbf{124}, no.13, 132002 (2020)
doi:10.1103/PhysRevLett.124.132002
[arXiv:1911.08123 [hep-lat]].

\bibitem{Colangelo:2014qya}
G.~Colangelo, M.~Hoferichter, A.~Nyffeler, M.~Passera and P.~Stoffer,
Phys. Lett. B \textbf{735}, 90-91 (2014)
doi:10.1016/j.physletb.2014.06.012
[arXiv:1403.7512 [hep-ph]].


\bibitem{Davier:2017zfy}
M.~Davier, A.~Hoecker, B.~Malaescu and Z.~Zhang,
Eur. Phys. J. C \textbf{77}, no.12, 827 (2017)
doi:10.1140/epjc/s10052-017-5161-6
[arXiv:1706.09436 [hep-ph]].

\bibitem{Davier:2019can}
M.~Davier, A.~Hoecker, B.~Malaescu and Z.~Zhang,
Eur. Phys. J. C \textbf{80}, no.3, 241 (2020)
[erratum: Eur. Phys. J. C \textbf{80}, no.5, 410 (2020)]
doi:10.1140/epjc/s10052-020-7792-2
[arXiv:1908.00921 [hep-ph]].



\bibitem{Borsanyi:2020mff}
S.~Borsanyi, Z.~Fodor, J.~N.~Guenther, C.~Hoelbling, S.~D.~Katz, L.~Lellouch, T.~Lippert, K.~Miura, L.~Parato and K.~K.~Szabo, \textit{et al.}
Nature \textbf{593} (2021) no.7857, 51-55
doi:10.1038/s41586-021-03418-1
[arXiv:2002.12347 [hep-lat]].

\bibitem{Alexandrou:2022amy}
C.~Alexandrou, S.~Bacchio, P.~Dimopoulos, J.~Finkenrath, R.~Frezzotti, G.~Gagliardi, M.~Garofalo, K.~Hadjiyiannakou, B.~Kostrzewa and K.~Jansen, \textit{et al.}
[arXiv:2206.15084 [hep-lat]].

\bibitem{Ce:2022kxy}
M.~C\`e, A.~G\'erardin, G.~von Hippel, R.~J.~Hudspith, S.~Kuberski, H.~B.~Meyer, K.~Miura, D.~Mohler, K.~Ottnad and P.~Srijit, \textit{et al.}
[arXiv:2206.06582 [hep-lat]].


  
\bibitem{Crivellin:2020zul} 
  A.~Crivellin, M.~Hoferichter, C.~A.~Manzari and M.~Montull,
  arXiv:2003.04886 [hep-ph].
  
\bibitem{deRafael:2020uif}
E.~de Rafael,
Phys. Rev. D \textbf{102} (2020) no.5, 056025
[arXiv:2006.13880 [hep-ph]].

\bibitem{Keshavarzi:2020bfy}
A.~Keshavarzi, W.~J.~Marciano, M.~Passera and A.~Sirlin,
Phys. Rev. D \textbf{102} (2020) no.3, 033002
doi:10.1103/PhysRevD.102.033002
[arXiv:2006.12666 [hep-ph]].

\bibitem{CMD-3:2023alj}
F.~V.~Ignatov \textit{et al.} [CMD-3],
Phys. Rev. D \textbf{109} (2024) no.11, 112002
doi:10.1103/PhysRevD.109.112002
[arXiv:2302.08834 [hep-ex]].
  
\bibitem{Passera:2008jk}
M.~Passera, W.~J.~Marciano and A.~Sirlin,
Phys. Rev. D \textbf{78} (2008), 013009
[arXiv:0804.1142 [hep-ph]].


   
\bibitem{Altmannshofer:2014pba}
W.~Altmannshofer, S.~Gori, M.~Pospelov and I.~Yavin,
Phys. Rev. Lett. \textbf{113} (2014), 091801
doi:10.1103/PhysRevLett.113.091801
[arXiv:1406.2332 [hep-ph]].

\bibitem{Athron:2021iuf}
P.~Athron, C.~Bal\'azs, D.~H.~J.~Jacob, W.~Kotlarski, D.~St\"ockinger and H.~St\"ockinger-Kim,
JHEP \textbf{09} (2021), 080
doi:10.1007/JHEP09(2021)080
[arXiv:2104.03691 [hep-ph]].


\bibitem{Lindner:2016bgg} 
  M.~Lindner, M.~Platscher and F.~S.~Queiroz,
  Phys.\ Rept.\  {\bf 731}, 1 (2018)
  [arXiv:1610.06587 [hep-ph]].

\bibitem{Crivellin:2021rbq}
A.~Crivellin and M.~Hoferichter,
JHEP \textbf{07} (2021), 135
doi:10.1007/JHEP07(2021)135
[arXiv:2104.03202 [hep-ph]].


\bibitem{Guedes:2022cfy}
G.~Guedes and P.~Olgoso,
[arXiv:2205.04480 [hep-ph]].




\bibitem{Baek:2016kud} 
  S.~Baek, T.~Nomura and H.~Okada,
  Phys.\ Lett.\ B {\bf 759}, 91 (2016)
  [arXiv:1604.03738 [hep-ph]].

\bibitem{Nomura:2018lsx}
T.~Nomura and H.~Okada,
Phys. Dark Univ. \textbf{26} (2019), 100359
doi:10.1016/j.dark.2019.100359
[arXiv:1808.05476 [hep-ph]].


\bibitem{Nomura:2018cfu}
T.~Nomura and H.~Okada,
Phys. Rev. D \textbf{99} (2019) no.5, 055027
doi:10.1103/PhysRevD.99.055027
[arXiv:1807.04555 [hep-ph]].

\bibitem{Anamiati:2018cuq}
G.~Anamiati, O.~Castillo-Felisola, R.~M.~Fonseca, J.~C.~Helo and M.~Hirsch,
JHEP \textbf{12} (2018), 066
doi:10.1007/JHEP12(2018)066
[arXiv:1806.07264 [hep-ph]].



\bibitem{Nomura:2018ibs}
T.~Nomura and H.~Okada,
Phys. Rev. D \textbf{99} (2019) no.5, 055033
doi:10.1103/PhysRevD.99.055033
[arXiv:1806.07182 [hep-ph]].


\bibitem{Nomura:2018cle} 
  T.~Nomura and H.~Okada,
  Phys.\ Lett.\ B {\bf 783}, 381 (2018)
  [arXiv:1805.03942 [hep-ph]].
  
\bibitem{Calibbi:2018rzv}
L.~Calibbi, R.~Ziegler and J.~Zupan,
JHEP \textbf{07} (2018), 046
doi:10.1007/JHEP07(2018)046
[arXiv:1804.00009 [hep-ph]].

  
\bibitem{Nomura:2017abu} 
  T.~Nomura and H.~Okada,
  Phys.\ Rev.\ D {\bf 96}, no. 9, 095017 (2017)
  [arXiv:1708.03204 [hep-ph]].
  
\bibitem{Cai:2017jrq}
Y.~Cai, J.~Herrero-Garc\'\i{}a, M.~A.~Schmidt, A.~Vicente and R.~R.~Volkas,
Front. in Phys. \textbf{5} (2017), 63
doi:10.3389/fphy.2017.00063
[arXiv:1706.08524 [hep-ph]].

  
\bibitem{Nomura:2016jnl} 
  T.~Nomura, H.~Okada and Y.~Orikasa,
  Phys.\ Rev.\ D {\bf 94}, no. 5, 055012 (2016)
  [arXiv:1605.02601 [hep-ph]].
  
  
\bibitem{Nomura:2016dnf} 
  T.~Nomura, H.~Okada and Y.~Orikasa,
  Phys.\ Rev.\ D {\bf 94}, no. 11, 115018 (2016)
  [arXiv:1610.04729 [hep-ph]].
  
\bibitem{Chen:2020jvl}
C.~H.~Chen and T.~Nomura,
Nucl. Phys. B \textbf{964} (2021), 115314
[arXiv:2003.07638 [hep-ph]].


\bibitem{Magg:1980ut} 
 M.~Magg and C.~Wetterich,
  Phys.\ Lett.\ B {\bf 94}, 61 (1980).
  
   \bibitem{Lazarides:1980nt} 
  G.~Lazarides, Q.~Shafi and C.~Wetterich,
  Nucl.\ Phys.\ B {\bf 181}, 287 (1981).
  
  
   \bibitem{Schechter:1980gr} 
  J.~Schechter and J.~W.~F.~Valle,
  Phys.\ Rev.\ D {\bf 22}, 2227 (1980).
  
\bibitem{Cheng:1980qt} 
  T.~P.~Cheng and L.~-F.~Li,
  Phys.\ Rev.\ D {\bf 22}, 2860 (1980).
  
  \bibitem{Mohapatra:1980yp} 
  R.~N.~Mohapatra and G.~Senjanovic,
  Phys.\ Rev.\ D {\bf 23}, 165 (1981).

 \bibitem{Bilenky:1980cx} 
  S.~M.~Bilenky, J.~Hosek and S.~T.~Petcov,
  Phys.\ Lett.\ B {\bf 94}, 495 (1980).



\bibitem{ParticleDataGroup:2020ssz}
P.~A.~Zyla \textit{et al.} [Particle Data Group],
PTEP \textbf{2020} (2020) no.8, 083C01
doi:10.1093/ptep/ptaa104




\bibitem{Cirelli:2005uq} 
  M.~Cirelli, N.~Fornengo and A.~Strumia,
  Nucl.\ Phys.\ B {\bf 753}, 178 (2006)
  [hep-ph/0512090].


\bibitem{Peskin:1990zt}
M.~E.~Peskin and T.~Takeuchi,
Phys. Rev. Lett. \textbf{65} (1990), 964-967
doi:10.1103/PhysRevLett.65.964




\bibitem{MEG:2016leq}
A.~M.~Baldini \textit{et al.} [MEG],
Eur. Phys. J. C \textbf{76} (2016) no.8, 434
doi:10.1140/epjc/s10052-016-4271-x
[arXiv:1605.05081 [hep-ex]].

   
\bibitem{MEG:2013oxv}
J.~Adam \textit{et al.} [MEG],
Phys. Rev. Lett. \textbf{110} (2013), 201801
doi:10.1103/PhysRevLett.110.201801
[arXiv:1303.0754 [hep-ex]].




  
      \bibitem{Belyaev:2012qa} 
  A.~Belyaev, N.~D.~Christensen and A.~Pukhov,
  Comput.\ Phys.\ Commun.\  {\bf 184}, 1729 (2013)
  [arXiv:1207.6082 [hep-ph]].
  
\bibitem{Nadolsky:2008zw} 
  P.~M.~Nadolsky, H.~L.~Lai, Q.~H.~Cao, J.~Huston, J.~Pumplin, D.~Stump, W.~K.~Tung and C.-P.~Yuan,
  Phys.\ Rev.\ D {\bf 78}, 013004 (2008)
  [arXiv:0802.0007 [hep-ph]].

\bibitem{FCC:2018byv}
A.~Abada \textit{et al.} [FCC],
Eur. Phys. J. C \textbf{79} (2019) no.6, 474
doi:10.1140/epjc/s10052-019-6904-3

\bibitem{Alwall:2014hca} 
  J.~Alwall {\it et al.},
  JHEP {\bf 1407}, 079 (2014)
  [arXiv:1405.0301 [hep-ph]].
  
  \bibitem{Alloul:2013bka} 
  A.~Alloul, N.~D.~Christensen, C.~Degrande, C.~Duhr and B.~Fuks,
  Comput.\ Phys.\ Commun.\  {\bf 185}, 2250 (2014)
  [arXiv:1310.1921 [hep-ph]].
  
\bibitem{Sjostrand:2014zea}
T.~Sj\"ostrand, S.~Ask, J.~R.~Christiansen, R.~Corke, N.~Desai, P.~Ilten, S.~Mrenna, S.~Prestel, C.~O.~Rasmussen and P.~Z.~Skands,
Comput. Phys. Commun. \textbf{191} (2015), 159-177
[arXiv:1410.3012 [hep-ph]].

\bibitem{deFavereau:2013fsa}
J.~de Favereau \textit{et al.} [DELPHES 3],
JHEP \textbf{02} (2014), 057
[arXiv:1307.6346 [hep-ex]].

   
\end{thebibliography}
\end{document}